\documentclass{ifacconf}
\usepackage{natbib}

\usepackage{amsmath,amssymb,mathtools}

\usepackage{graphicx}
\usepackage{float}

\usepackage{algorithm}
\usepackage{algorithmic}
\usepackage{multirow} 
\usepackage[table]{xcolor}
\usepackage{array}
\usepackage{tabularx}
\usepackage{caption}

\newcolumntype{L}[1]{>{\raggedright\arraybackslash}p{#1}}
\newcolumntype{C}[1]{>{\centering\arraybackslash}p{#1}}

\begin{document}
\begin{frontmatter}
\title{Comparing Point and Interval Methods for Equilibrium Computation under Parametric Uncertainty}
\author[First]{R. Prakash} 
\author[First]{S. Janardhanan} 
\author[First]{S. Sen}
\address[First]{Department of Electrical Engineering, Indian Institute of Technology Delhi, New Delhi--110016, India (e-mail: rudra.prakash@ee.iitd.ac.in, janas@ee.iitd.ac.in, shaunak.sen@ee.iitd.ac.in).}

\begin{abstract}
Equilibrium points define operating conditions for nonlinear dynamical and control systems. Their existence, multiplicity, and stability under parametric uncertainty determine feasible operating regimes and the validity of robustness claims. With parameters constrained to a bounded set, one can (i) compute equilibria at sampled parameter values, (ii) trace equilibria along a prescribed path in parameter space, or (iii) identify states in a given operating domain that are equilibria for at least one admissible parameter realization. We compare standard pointwise workflows—direct simulation, numerical continuation, residual minimization, and a multistart Newton–Raphson method—with validated interval-analysis–based workflows. The latter (a) provide formal certificates of exclusion, existence, and uniqueness of equilibria for fixed parameters and, under parametric inclusion conditions, uniformly over entire parameter boxes, and (b) construct rigorous outer enclosures in state space that provably contain all equilibria associated with the full admissible parameter set. Biomolecular circuit models governed by nonlinear ODEs serve as a representative application domain. We benchmark three canonical architectures across four levels of parameter uncertainty, including a genetic toggle switch near a symmetry-breaking bifurcation. Sampling- and slice-based approaches can miss or underrepresent multistability, whereas interval-based outer enclosures yield mathematically rigorous bounds on the equilibrium set induced by parametric uncertainty.
\end{abstract}


\end{frontmatter}

\section{Introduction}
Equilibria serve as operating points for nonlinear dynamical and control systems; their multiplicity and local stability determine admissible modes of operation and underpin stability analysis and controller validation.
Under parametric uncertainty, whether an operating point exists, is unique, or admits multiple stable equilibria directly affects robustness assessments and the characterization of operating envelopes.
Biomolecular circuits constitute a representative application setting in which these issues are pronounced, as they are commonly modeled as nonlinear ODEs with rational or Hill-type nonlinearities~\citep{del_vecchio_biomolecular_2015,alon_introduction_2006} and can exhibit switching, homeostasis, and adaptation.
Kinetic parameters are uncertain due to biological variability, context dependence, and limited identifiability; consequently, analysis is often conducted over an admissible set $P$ rather than a single nominal parameter vector $p\in P$.
Under bounded uncertainty, equilibrium analysis can target distinct objectives: computing equilibria at sampled parameter instances, tracing equilibria along a chosen low-dimensional slice of $P$, or characterizing regions of state space that contain equilibria for at least one admissible parameter.
To facilitate worst-case analysis over a prescribed operating domain $X$, we consider the existential equilibrium projection
\(
\mathcal{S}_X(P)=\{x\in X\mid \exists p\in P:\ f(x,p)=0\},
\)
which denotes the set of all states that constitute equilibria for at least one admissible parameter value.

\begin{figure}[h!]
\centering
\includegraphics[width=0.75\linewidth]{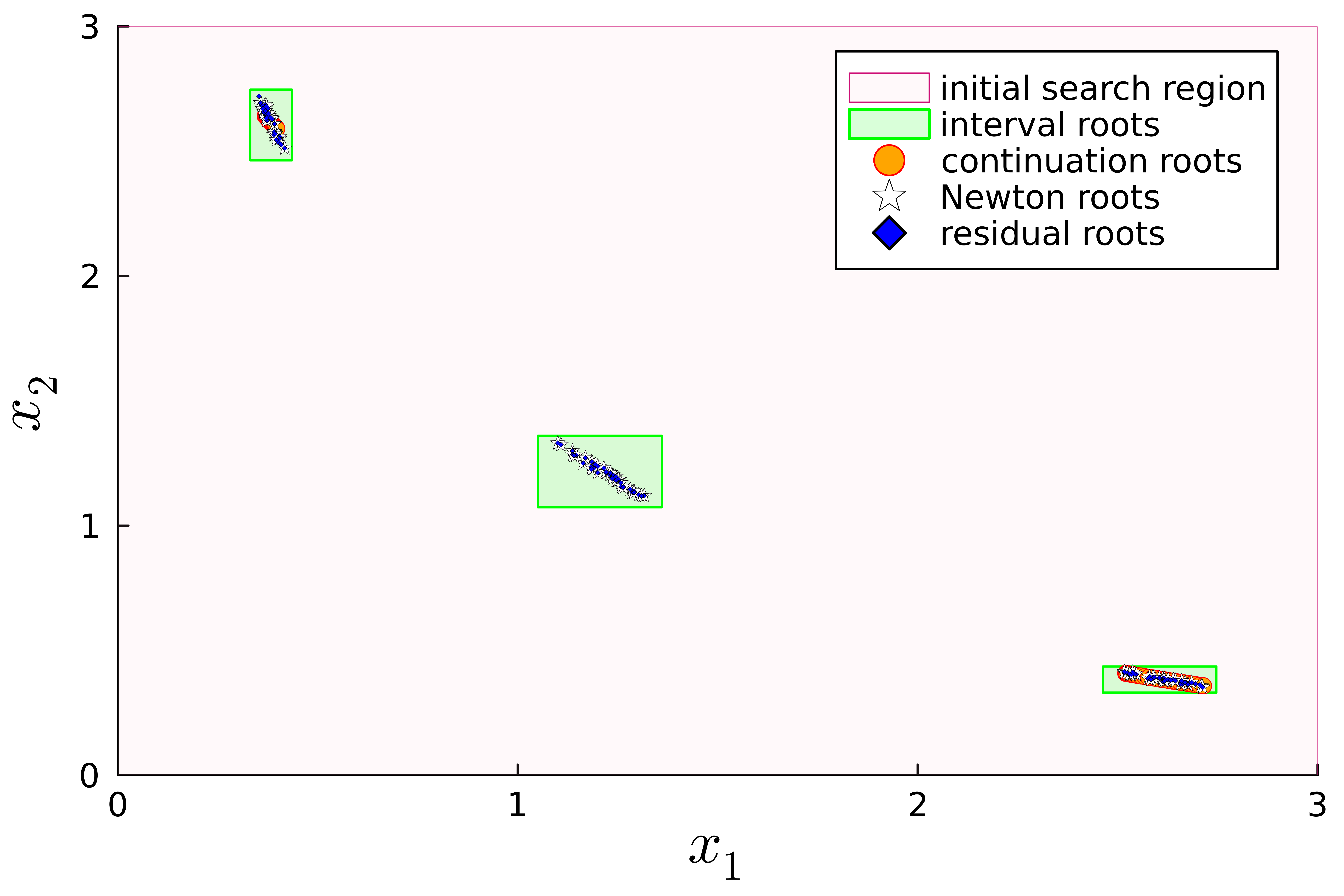}
\caption{Overlay of point-based equilibrium candidates and an interval outer enclosure for the uncertain toggle-switch at \(\rho=0.02\); enclosure computed on \(X=[0,3]^2\)). Pink box: initial search region. Green boxes: outer enclosure of roots for \(P_{\rho}=[2.94,3.06]\times[2.94,3.06]\). Orange filled circles: stable continuation equilibria along \(\alpha_2=3\), \(\alpha_1\in[2.94,3.06]\). White star: accepted multistart Newton roots for parameters sampled in \(P_{\rho}\). Blue diamond are residual roots.}
\label{fig:toggle_point_interval_overlay}
\end{figure}

We compare five equilibrium-computation workflows---simulation, continuation, residual minimization, multistart Newton, and interval analysis---and evaluate them on three benchmark circuits across four uncertainty widths (Table~\ref{tab:comparison}).
Simulation, continuation, residual minimization, and multistart Newton provide candidate equilibria, but the supported inferences depend on sampling density, initialization, and slice selection (e.g.,~\citep{Kuznetsov2004}).
Interval methods complement these strategies by (i) rigorously certifying exclusion, existence, and uniqueness for specific parameter instances (and, when parametric inclusion conditions hold, uniformly over an entire parameter box) and (ii) computing guaranteed outer enclosures of $\mathcal{S}_X(P)$ for uncertain parameter sets, with computational cost that can grow with both the state dimension and the size of the uncertainty region~\citep{neumaier_interval_1990,KearfottRigorousGlobalSearch}.
Related studies apply interval techniques to steady-state uncertainty quantification in biomolecular networks~\citep{chorasiya_quantitative_2023,prakash_rigorous_2025} and examine computational scaling in validated pipelines~\citep{prakash2026computational,tucker_validated_2011}.
A genetic toggle switch~\citep{Gardner2000} operated near a symmetry-breaking bifurcation highlights that sampling- and slice-based analyses can miss or underrepresent multistability, whereas interval-based outer enclosures yield conservative worst-case localization (Figure~\ref{fig:toggle_point_interval_overlay}).
These differences matter when equilibrium computations underpin robust operating-regime claims in feedback design and verification under parametric variability.

The main contributions are: (i) We formally specify objectives for equilibrium computation under bounded parametric uncertainty via explicit quantification over $p \in P$.
(ii) We systematically connect five widely used computational workflows---simulation, numerical continuation, residual minimization, multistart Newton--Raphson, and interval analysis---to the uncertainty statements they can substantiate, making explicit the guarantees associated with their typical outputs.
(iii) We propose a reproducible, claim-focused reporting template (Table~\ref{tab:claim_oriented_summary}) that separates (a) sample-based or slice-based numerical evidence, (b) validated certificates for fixed parameter values $p$, and (c) interval-derived outer enclosures for uncertain parameter sets $P$.
(iv) Through benchmarking, we demonstrate how point-based and validation-based workflows can lead to different inferences near multistability and bifurcations, and we distill practical consequences for robust analysis.

The contribution is primarily conceptual and evaluative: we determine which types of uncertainty claims are supported by standard computational workflows and design an empirical protocol to rigorously stress-test those claims.
The comparison does not produce a single best method; instead, it reveals that sampling-based, slice-based, and rigorously validated set-based computations address distinct inferential questions, so conclusions must employ quantifiers aligned with the epistemic guarantees characteristic of each method family.

Section~\ref{sec:problem} formulates the problem; Section~\ref{sec:methods} details the computational workflows; Section~\ref{sec:comparison} analyzes theoretical guarantees and scaling behavior; Section~\ref{sec:case} reports the computational experiments; and Section~\ref{sec:discussion} provides implementation-oriented guidance.
\section{Problem formulation}\label{sec:problem}
We consider an autonomous nonlinear dynamical system of the form
\begin{equation}
\dot{x} = f(x,p), \qquad x \in \mathbb{R}^n,\quad p \in P \subset \mathbb{R}^m,
\label{eq:ode}
\end{equation}
where $f$ is continuously differentiable with respect to $x$ on the state domain of interest and $P$ denotes a compact set of admissible parameter values.
In the computational study, $P$ is specified as an axis-aligned interval box.

For a fixed parameter realization $p \in P$, an equilibrium $x^\star$ is defined as a solution of
\(
 f(x^\star,p)=0.
 \label{eq:ss_fixed}
\)
When $p$ is allowed to vary over the admissible parameter set $P$, different computational workflows implicitly correspond to different uncertainty-handling objectives.

A central set-valued construct for worst-case analysis is the existential equilibrium projection on a prescribed state domain $X \subset \mathbb{R}^n$, given by
\begin{equation}
\mathcal{S}_X(P) := \{x \in X \mid \exists\, p \in P:\ f(x,p)=0\}.
\label{eq:SX_exist}
\end{equation}
Interval outer-enclosure techniques aim to compute a mathematically rigorous superset of \eqref{eq:SX_exist}. In contrast, sampling-based and slice-based approaches only yield pointwise evidence at selected parameter values $p$.

For a fixed parameter $p$, an equilibrium point $x^\star$ is said to be \emph{hyperbolic} if the Jacobian matrix $A := \partial f/\partial x (x^\star,p)$ has no eigenvalues lying on the imaginary axis. The equilibrium is (locally) exponentially stable if all eigenvalues satisfy $\Re(\lambda) < 0$, and it is unstable if there exists at least one eigenvalue with $\Re(\lambda) > 0$~\citep{KhalilNonlinearSystems}. 
In numerical experiments, long-time forward simulations tend to preferentially detect attracting invariant sets; in contrast, unstable equilibria typically necessitate the use of equation-based numerical solvers or techniques for rigorous (validated) certification.
\section{Computational methods}\label{sec:methods}
We present a systematic overview of computational frameworks for determining equilibria of~\eqref{eq:ode} in the presence of parametric uncertainty and delineate, for each workflow, the class of analytical or inferential statements that it rigorously supports.

\subsection{Point-based workflows (nonvalidated)}
\subsubsection{Simulation (sampling and sweeps).}
We sample or discretize the parameter space $P$ and numerically integrate~\eqref{eq:ode} (typically from multiple initial conditions) until a prescribed terminal residual criterion is met; the resulting terminal state is then interpreted as evidence of an attracting equilibrium.
This procedure exhibits favorable scaling with respect to the dimension of the parameter space, but it is systematically biased by the structure of basins of attraction and by the choice of stopping criteria, and it only rarely identifies unstable equilibria.
All convergence criteria and numerical solver configurations are summarized in \ref{subsec:numerical_settings}.

\subsubsection{Continuation and point solvers.}
Equation-based workflows solve the nonlinear system $f(x,p)=0$ directly.
Classical continuation methods track equilibria along a chosen parameter path and can follow both stable and unstable branches, but they are inherently local and may miss disconnected solution components unless supported by systematic seeding and explicit branch-management strategies~\citep{AllgowerGeorg}. More general continuation frameworks can handle multiple parameters, yet they remain path-based and thus require carefully designed exploration strategies to approximate the solution structure over a high-dimensional parameter space $P$.
By contrast, we use a multistart (random-restart) Newton method, i.e., Newton's method initialized from multiple starting points, to solve $f(x,p)=0$ independently at sampled parameter values. This procedure returns candidate roots, is sensitive to initialization, and provides no inherent validation or completeness guarantees.

\subsubsection{Residual minimization.}
An alternative approach consists in minimizing the squared residual over a bounded domain,
\begin{equation}
\min_{x\in X}\ \phi(x;p):=\tfrac12\|f(x,p)\|_2^2,
\label{eq:lsq}
\end{equation}
which naturally incorporates constraints but yields a generally nonconvex optimization problem and may converge to stationary points that do not correspond to roots of the underlying system~\citep{DennisSchnabel}.
\subsection{Interval workflows (validated)}
Interval analysis replaces pointwise evaluations with guaranteed enclosures over multidimensional intervals (``boxes''), enabling validated exclusion tests and, for fixed parameters $p$, local existence and uniqueness certification of equilibria.
A contract-and-subdivide search on $X$ recursively bisects boxes and applies interval Newton and Krawczyk tests~\citep{MooreInterval,neumaier_interval_1990,jaulin_applied_2001}:
(i) if $0\notin f(B,p)$, then $B$ contains no equilibrium for that $p$;
(ii) if standard Krawczyk/Newton contraction conditions hold, $B$ is certified to contain a unique equilibrium.

For uncertain parameters, two validated goals arise. First, a \emph{parametric} interval-Newton/Krawczyk test can, under usual regularity assumptions on $\partial f/\partial x$ over $X\times P$, certify the uniform statement $\forall p\in P,\ \exists!\,x^*(p)\in X$ with $f(x^*(p),p)=0$.
Second, for existential worst-case localization on a given operating domain, we use interval range exclusion on $f(B,P)$ to discard boxes that cannot contain equilibria for any $p\in P$, and keep the rest as a rigorous outer enclosure of \eqref{eq:SX_exist}.
This enclosure is rigorous but generally inconclusive: retained boxes are neither guaranteed to contain an equilibrium nor to contain it uniquely.
Stopping criteria and computational budgets are given in \ref{subsec:numerical_settings}; outcomes appear as fixed-$p$ certifications and, for uncertain $P$, as undecided volumes and box counts in Table~\ref{tab:claim_oriented_summary}.
\section{Theoretical comparison}
\label{sec:comparison}
The empirical analysis emphasizes that, under bounded parametric uncertainty, the workflow classes differ not only in computational cost and output format, but also in the \emph{logical strength} of the statements they can validate.

The interval-based workflow separates two tasks that are often conflated. For a \emph{fixed} parameter instance $p$, Krawczyk or interval-Newton methods with domain subdivision can certify both exclusion and local existence/uniqueness of equilibria on given state-space boxes.
For an \emph{uncertain} parameter box $P$, there are two distinct types of validated statements:
(i) A parametric interval-Newton/Krawczyk inclusion test can, under standard regularity assumptions, certify \emph{uniform} existence and uniqueness, i.e., for every $p \in P$ there is exactly one root in $X$.
(ii) Interval range exclusion applied to $f(B,P)$ yields a rigorous outer enclosure of the existential projection $\mathcal{S}_X(P)$ in~\eqref{eq:SX_exist}: if $0 \notin f(B,P)$ for a state box $B \subseteq X$, then no state in $B$ is an equilibrium for any $p \in P$.
In this study, all results for uncertain sets $P$ are of type (ii): they provide worst-case exclusion and outer localization of equilibria, but no uniform-in-$P$ existence or uniqueness guarantees.
This distinction appears clearly in the numerical results. In the antithetic integral feedback benchmark at \(\rho = 0.20\), fixed-\(p\) certification typically confirms a single equilibrium for most representative parameter realizations (Table~\ref{tab:claim_oriented_summary}), whereas the uncertain-\(P\) outer enclosure over the full set \(P_\rho\) still has a nonzero undecided volume.
\subsection{Comparison of workflow classes}
Table~\ref{tab:comparison} summarizes the capabilities and limitations of the workflow classes used here. The goal is not to identify a universally superior method, but to align each workflow with the type of uncertainty statement it can rigorously support, thereby reducing the risk of over-interpreting the resulting inferences.
\begin{table*}[t]
\centering
\caption{Comparison of equilibrium-computation workflows under parameter uncertainty.}
\label{tab:comparison}
\footnotesize
\setlength{\tabcolsep}{4.5pt}
\renewcommand{\arraystretch}{1.2}
\begin{tabular}{|>{\raggedright\arraybackslash}m{1.5cm}|>{\raggedright\arraybackslash}m{2.1cm}|>{\raggedright\arraybackslash}m{2.15cm}|>{\raggedright\arraybackslash}m{2.5cm}|>{\raggedright\arraybackslash}m{2.2cm}|>{\raggedright\arraybackslash}m{5.8cm}|}
\hline
\rowcolor{blue!5}
\textbf{Aspect} & \textbf{Simulation} & \textbf{Continuation} & \textbf{Residual minimization} & \textbf{Multistart Newton} & \textbf{Interval analysis} \\
\hline
Numerical object & Terminal states from time simulation & Roots traced along a parameter path & Roots as minimizers of $\tfrac12\|f(x,p)\|_2^2$ & Roots of $f(x,p)=0$ from many starts & Interval boxes in $X$ (fixed $p$) or $X\times P$ (uncertain $P$): excluded / certified / retained \\
\hline
Supported statement & Empirical evidence of attracting equilibria (no completeness) & Local branch/slice evidence & Candidate roots only; small residual $\not\Rightarrow f=0$ & Candidate roots at sampled $p$ (no completeness) & Fixed $p$: certify exclusion, existence, and uniqueness (per box). Uncertain $P$: compute a rigorous outer enclosure of $S_X(P)$; under parametric inclusion conditions, also certify $\forall p\in P,\ \exists!\,x^*(p)\in X$. \\
\hline
Unstable equilibria & Typically not recovered & Recovered on traced branches & Possible; depends on initialization & Possible; depends on initialization & Can rigorously calculate if exists for some \(p \in P\).\\
\hline
Uncertainty handling & Sample $p\in P$ & Sweep 1--few parameters in $P$ & Solve at sampled $p$ & Solve at sampled $p$ & Interval contractors act on $X$ or $X\times P$ \\
\hline
Main limitation / best use & Basin dependence; computationally efficient screening & Path dependence & Prone to non-root local minima; sensitive to scaling and initialization & No validation; sensitive to singular Jacobians & Conservative; subdivision cost; appropriate for validated screening and certification \\
\hline
\end{tabular}
\normalsize
\end{table*}
An equilibrium point \(x^\star\) is locally asymptotically stable for a fixed parameter value \(p\) if the Jacobian matrix \(J(x^\star,p)=\partial f/\partial x\) evaluated at \((x^\star,p)\) has eigenvalues whose real parts are all strictly negative.
Consequently, trajectory-based simulations predominantly reveal attracting equilibria, whereas equation-based workflows are also capable of locating unstable equilibria when the numerical search is initialized in, or reaches, their neighborhoods.
For instance, in the positive-autoregulation benchmark, numerical continuation identifies both stable and unstable equilibria along each parameter slice, while time-domain simulations yield only attracting terminal states.
The reported continuation counts depend on the specific nearest-neighbor matching scheme employed to associate roots between adjacent slices and should therefore be interpreted as counts of detected continuation segments, rather than as exact topological branch counts.
\subsubsection{Scaling and observed resolution trends.}
Sampling-based simulation complexity scales with the number of parameter realizations, initial conditions per realization, and the cost of forward time integration.  
Continuation methods scale with the number of parameter sweep points and the cost of local corrections at each discretized point on the continuation curve.  
Residual-based optimization scales with the number of restarts and the cost of local descent on a nonconvex objective.  
Interval analysis scales with the size of the subdivision tree over the state space \(X\), which can grow exponentially with state dimension and is highly sensitive to interval overestimation~\citep{neumaier_interval_1990,jaulin_applied_2001,HansenWalster2004,MooreKearfottCloud2009,prakash2026computational}.

For interval outer enclosures with uncertain parameters, computational complexity is well indicated by the pair (residual undecided volume, total number of processed boxes). Using the results in Section~\ref{sec:case} (Table~\ref{tab:claim_oriented_summary}) as examples, the toggle-switch benchmark increases from \((0.0017, 5379)\) at \(\rho = 0.02\) to \((0.0962, 207221)\) at \(\rho = 0.20\); the positive-autoregulation benchmark from \((0.0322, 225)\) to \((0.2070, 579)\); and the antithetic integral-feedback benchmark from \((1.3\times 10^{-9}, 402901)\) to \((1.6\times 10^{-7}, 46086279)\).

In the case studies, point-based workflows produce evidence that depends on sampling strategies and local search heuristics, whereas interval-based workflows provide mathematically certified exclusion and conservative localization, with resolution limited by the subdivision depth and interval overestimation.
\section{Case studies}
\label{sec:case}
\subsection{Evaluation protocol}
We compare five workflow classes across three benchmark models and four relative uncertainty widths \(\rho \in \{0.02,\,0.05,\,0.10,\,0.20\}\). For each benchmark, the uncertainty domain is obtained by applying the relative perturbation \(\rho\) to the specified coordinates of the nominal parameter vector. All other settings are fixed across uncertainty widths: in simulation, we use $40$ parameter samples and $16$ initial conditions per sample; in continuation, we run forward and backward sweeps on a $61$-point grid ($122$ directional slices); in residual minimization, we use $40$ parameter samples and $24$ starting points; in multistart Newton, $40$ parameter samples and $16$ starting points; and in interval analysis, we combine fixed-$p$ certification on representative parameter instances with an uncertain-$P$ outer enclosure over \(P_\rho\). The numbers of representative fixed-$p$ instances are $7$ for the toggle switch, $3$ for the positive autoregulation circuit, and $11$ for the antithetic integral feedback system. Numerical tolerances and computational budgets are given in \ref{subsec:numerical_settings}.

Given a prescribed operating region $X$ and an uncertainty set $P_\rho$, a central robustness question is whether additional equilibrium points can arise within $X$ for any admissible parameter realization. The uncertain-$P$ interval workflow addresses this question by constructing a mathematically rigorous outer enclosure of $\mathcal{S}_X(P_\rho)$, which is quantified through the undecided volume and the collection of processed boxes. In contrast, numerical continuation and multistart Newton yield candidate evidence only on lower-dimensional slices of $P_\rho$ or at selected sampled parameter values. By design, continuation is slice-based and does not constitute a global statement over the entire uncertainty domain $P_\rho$. Fixed-$p$ certification augments these candidate computations by rigorously validating equilibrium root counts at representative parameter instances.
\subsection{Numerical settings}\label{subsec:numerical_settings}
Simulation uses relative and absolute tolerances of $10^{-8}$ and declares convergence when the tail-window residual satisfies $\max_{t\in\mathcal{T}_{\mathrm{tail}}}\|f(x(t),p)\|_2\le 10^{-6}$ (with horizons $T=300$ for the toggle switch and positive autoregulation benchmarks and $T=1000$ for the antithetic benchmark). Continuation and multistart Newton accept roots when $\|f(x,p)\|_2\le 10^{-10}$, whereas residual minimization accepts candidates after polishing when $\|f(x,p)\|_2\le 10^{-6}$. Interval analysis uses widest-coordinate bisection with midpoint-Jacobian Krawczyk tests and a resolution factor $10^{-3}$, with no maximum depth or maximum boxes limits imposed.
We report workflow-specific performance indicators as follows: convergence fractions and terminal cluster counts for the simulation workflow; numbers of detected continuation segments and associated stability tallies for the continuation workflow; root-hit fractions and counts of stationary non-root points for the residual minimization workflow; and excluded as well as undecided volume fractions for the interval analysis workflow. The undecided volume fraction is defined as the total volume of undecided boxes normalized by $\operatorname{vol}(X)$.
\subsection{Genetic toggle switch}
The toggle-switch system serves as the analytically tractable reference benchmark in this study and is based on the canonical genetic toggle-switch architecture~\citep{Gardner2000}. The governing dynamical model is
\begin{equation}
\label{eq:toggle}
\begin{aligned}
\dot{x}_1 &= \frac{\alpha_1}{1+x_2^{\beta}} - x_1, \qquad
\dot{x}_2 = \frac{\alpha_2}{1+x_1^{\gamma}} - x_2.
\end{aligned}
\end{equation}
Nominal parameter values are \((\alpha_1,\alpha_2,\beta,\gamma)=(3,3,2,2)\).
In the symmetric case \(\alpha_1=\alpha_2=\alpha\) with \(\beta=\gamma=2\), a symmetry-breaking bifurcation occurs at \(\alpha=2\). The nominal choice \(\alpha=3\) used in simulations is therefore in the bistable regime. In the parameter-uncertainty study, \(\alpha_1\) and \(\alpha_2\) vary independently around \(3\), while \(\beta\) and \(\gamma\) remain fixed at \(2\). For continuation, \(\alpha_1\) is the continuation parameter and is swept over its uncertainty range with \(\alpha_2\) fixed at \(3\). Thus, the continuation diagrams show the one-parameter slice \(\alpha_2=3\) of the full two-parameter uncertainty domain.
Table~\ref{tab:claim_oriented_summary} summarizes how the observed outcomes vary with uncertainty width across the three benchmark circuits studied: the toggle switch, the positive autoregulatory motif, and the antithetic integral feedback controller.
\begin{table*}[t]
\centering
\caption{Workflow-specific numerical summary (each column corresponds to a different uncertainty statement). \textbf{Simulation: convergence rate; clusters per $p$ (min/med/max):} ``$c;\, b_{\min}/b_{\mathrm{med}}/b_{\max}$'', where $c$ is the fraction of simulation runs passing the terminal residual test and $b_{\min}/b_{\mathrm{med}}/b_{\max}$ are the min/median/max counts of distinct attracting terminal-state clusters per sampled parameter. \textbf{Continuation: roots per slice:} root pattern along a one-parameter continuation slice; $S/U/H$ denote stable/unstable/marginal (nonhyperbolic) equilibria; ``$+$'' concatenates counts (e.g., $1S+1U$), and ``--'' indicates a range across slices. \textbf{Residual hit fraction $(10^{-6}/10^{-8})$:} fraction of residual-minimization runs accepted as roots at $\|f\|_2\le 10^{-6}$ and $10^{-8}$. \textbf{Multistart Newton roots/$p$:} min/median/max distinct accepted multistart Newton roots per sampled parameter. \textbf{Fixed-$p$ certificates (coverage; roots):} ``$m/n; r$'', where $m$ of $n$ representative parameters were fully certified and $r$ is the certified root count on those instances (possibly a range). \textbf{Uncertain-$P$ undecided volume:} undecided volume fraction of the uncertain-parameter interval outer enclosure over the full box $P_\rho$ (smaller is better). \textbf{Processed boxes:} total processed interval boxes in the uncertain-$P$ computation.}
\label{tab:claim_oriented_summary}
\footnotesize
\setlength{\tabcolsep}{4.5pt}
\renewcommand{\arraystretch}{1.2}
\begin{tabular}{|>{\raggedright\arraybackslash}m{1.71cm}|>{\centering\arraybackslash}m{0.65cm}|>{\raggedright\arraybackslash}m{2.35cm}|>{\raggedright\arraybackslash}m{2.05cm}|>{\raggedright\arraybackslash}m{1.7cm}|>{\raggedright\arraybackslash}m{2.3cm}|>{\raggedright\arraybackslash}m{1.45cm}|>{\raggedright\arraybackslash}m{1.55cm}|>{\raggedright\arraybackslash}m{1.35cm}|}
\hline
\rowcolor{blue!5}
\textbf{Benchmark} & $\boldsymbol{\rho}$ & \textbf{Simulation: convergence rate; clusters per $p$ (min/med/max)} & \textbf{Continuation: roots per slice} & \textbf{Residual hit fraction $(10^{-6}/10^{-8})$} & \textbf{Multistart Newton roots/$p$ (min/med/max)} & \textbf{Fixed-$p$ certificates (coverage; roots)} & \textbf{Uncertain-$P$ undecided volume} & \textbf{Processed boxes} \\
\hline
\multirow{4}{=}{Toggle switch} & 0.02 & 1.000; 2/2/2 & 2S+1U & 1.000 / 0.477 & 1/3/3 & 7/7; 3 & 0.0017 & 5379 \\
& 0.05 & 1.000; 2/2/2 & 2S+1U & 1.000 / 0.554 & 1/3/3 & 7/7; 3 & 0.0095 & 24143 \\
& 0.10 & 1.000; 1/2/2 & 0--2S+1U & 0.925 / 0.518 & 0/3/3 & 7/7; 1--3 & 0.0388 & 86899 \\
& 0.20 & 1.000; 1/1/2 & 0--2S+1U & 0.435 / 0.246 & 0/1/3 & 7/7; 0--3 & 0.0962 & 207221 \\
\hline
\multirow{4}{=}{Positive autoregulation} & 0.02 & 1.000; 2/2/2 & 2S+1U & 1.000 / 0.618 & 3/3/3 & 3/3; 3 & 0.0322 & 225 \\
& 0.05 & 1.000; 2/2/2 & 2S+1U & 1.000 / 0.627 & 3/3/3 & 3/3; 3 & 0.0605 & 285 \\
& 0.10 & 1.000; 2/2/2 & 2S+1U & 1.000 / 0.598 & 3/3/3 & 3/3; 3 & 0.1084 & 381 \\
& 0.20 & 1.000; 2/2/2 & 2S+1U & 1.000 / 0.514 & 3/3/3 & 3/3; 3 & 0.2070 & 579 \\
\hline
\multirow{4}{=}{Antithetic integral feedback} & 0.02 & 0.905; 0/1/1 & 1S & 1.000 / 0.410 & 1/1/1 & 11/11; 1 & $1.3\times 10^{-9}$ & 402901 \\
& 0.05 & 0.756; 0/1/1 & 1S & 1.000 / 0.376 & 1/1/1 & 11/11; 1 & $8.2\times 10^{-9}$ & 2424403 \\
& 0.10 & 0.652; 0/1/1 & 1S & 1.000 / 0.403 & 1/1/1 & 11/11; 1 & $3.4\times 10^{-8}$ & 9902509 \\
& 0.20 & 0.722; 0/1/1 & 1S & 1.000 / 0.356 & 1/1/1 & 11/11; 1 & $1.6\times 10^{-7}$ & 46086279 \\
\hline
\end{tabular}
\normalsize
\end{table*}
Table~\ref{tab:claim_oriented_summary} presents a workflow-specific summary, and its caption defines the column semantics and the shorthand notation used.
Figure~\ref{fig:toggle_point_interval_overlay} illustrates the relationship between the sampled point-based candidates and the uncertain-\(P\) interval outer enclosure at \(\rho = 0.02\), for the parameter domain \(P_{\rho} = [2.94, 3.06] \times [2.94, 3.06]\) and the continuation slice defined by \(\alpha_2 = 3\) and \(\alpha_1 \in [2.94, 3.06]\). Table~\ref{tab:claim_oriented_summary} presents a consolidated overview of the width-dependent behavior exhibited by the various workflows.
As \(\rho\) increases, simulations converge with higher frequency; however, continuation remains slice-based (varying solely \(\alpha_1\) with \(\alpha_2\) fixed) and may fail to detect equilibria that occur elsewhere in the two-parameter uncertainty set.
Interval analysis partitions the operating domain: the uncertain-\(P\) computation yields a rigorous outer enclosure (reported via the undecided volume and processed boxes), while fixed-$p$ certification on representative parameter instances can validate the number of equilibria on selected slices near the symmetry-breaking regime.
\subsection{Positive autoregulation}
The positive-autoregulation example is formulated as a one-dimensional Hill-type self-activation model, representing a canonical feedback motif in systems biology~\citep{alon_introduction_2006,del_vecchio_biomolecular_2015},
\begin{equation}
\dot{x}=\beta+\alpha\,\frac{x^h}{K^h+x^h}-\delta x,
\label{eq:autoreg}
\end{equation}
with nominal parameters $(\beta,\alpha,\delta, K, h)=(0.1,3.5,1.0,1.0,$ $4.0)$, parametric uncertainty applied to \(\alpha\), and operating domain \(X=[0,8]\). This benchmark isolates multistability in the minimal possible state dimension and is therefore well suited for systematically comparing residual-based root search with interval-based certification techniques.

In this one-dimensional setting, the pointwise workflows agree: direct simulation converges in all runs, numerical continuation recovers the expected stable and unstable equilibria, and multistart Newton finds one to three roots for each sampled parameter (Table~\ref{tab:claim_oriented_summary}). Fixed-\(p\) certification confirms three roots for all representative parameters, while the outer enclosure under uncertain \(P\) expands with \(\rho\), leaving a growing fraction of parameter space with undecided root existence.
\subsection{Antithetic integral feedback}
The antithetic integral feedback example 
is the most directly relevant to control-oriented biomolecular modeling, as it is derived from the antithetic integral feedback controller architecture~\citep{BriatGuptaKhammash2016}. The dynamical model is given by
\begin{equation}
\label{eq:antithetic}
\begin{aligned}
\dot{x}_1 &= \kappa z_1-\delta x_1, \quad
\dot{x}_2 = kx_1-\gamma x_2, \\
\dot{z}_1 &= \mu-\eta z_1z_2, \quad
\dot{z}_2 = \theta x_2-\eta z_1z_2,
\end{aligned}
\end{equation}
with nominal parameter values \((\delta,\gamma,k,\kappa,\mu,\theta,\eta)=(1,1,1,\) \(1,10,2,1)\), parametric uncertainty applied to \(\delta,\gamma,k,\kappa\).

This example contrasts empirical, simulation-based evidence with worst-case localization over the full uncertainty set (Table~\ref{tab:claim_oriented_summary}). Time-domain simulations converge only for a subset of trajectories and, even when convergent, primarily constrain the regulated output \(x_2\), leaving other state coordinates weakly localized.
Continuation on the prescribed slice and multistart Newton yield a single candidate equilibrium pattern at sampled parameter values. Interval analysis provides a set-based characterization: the uncertain-$P$ outer enclosure retains a nonzero undecided region, whereas fixed-$p$ certification can establish local uniqueness for representative parameters (e.g., at \(\rho=0.20\)); together, these results delineate the gap between numerical evidence and statements intended to hold uniformly over $P$.

Point-based results depend on sampling, initialization, and (for continuation) the chosen slice, so they need not reflect behavior over the full set $P$. For uncertain-$P$, our interval computations yield only outer enclosures of \eqref{eq:SX_exist}: discarded boxes rigorously certify absence of equilibria (for all $p\in P$), but retained boxes are undecided and do not certify existence or uniqueness. Tighter enclosures typically require model-specific scaling/preconditioning and, often, subdivision of $P$; a natural extension is hybrid candidate generation followed by validated certification and, when feasible, parametric inclusion tests for uniform-in-$P$ existence/uniqueness.
\section{Conclusions}
\label{sec:discussion}
This paper maps equilibrium-computation workflows to the uncertainty statements they support; Table~\ref{tab:comparison} and Table~\ref{tab:claim_oriented_summary} separate sampled/slice-based evidence from validated fixed-$p$ certificates and uncertain-$P$ outer enclosures.
Sampling-based simulation and multistart Newton can miss equilibria when basins are small or initializations are uninformative, while continuation is intrinsically slice-based and may omit solution components off the chosen path. 
Residual minimization may converge to non-root stationary points under poor scaling or near-singular Jacobians, and nearest-neighbor matching in continuation tallies detected path segments rather than global branch topology.
For fixed $p$, validated tests can certify exclusion/existence/uniqueness on state boxes; for uncertain $P$, interval range exclusion yields rigorous outer enclosures of the existential equilibrium projection, with the undecided volume quantifying residual ambiguity (a nonzero value indicates that additional refinement is required before making absence/presence claims over the full set).
Use point methods for candidate generation and qualitative exploration, but use validated fixed-$p$ certification to support root-count claims and uncertain-$P$ outer enclosures for worst-case localization over $P\times X$; when undecided volume remains large, tighten contractors and/or subdivide $P$ and $X$ before drawing robust conclusions, and report the intended quantifiers explicitly.
\bibliography{ifacconf} 
\end{document}